
 \message
 {JNL.TEX version 0.8 as of 10/28/85.  Report bugs and problems to Doug
Eardley.}



 \font\twelverm=cmr10 at 12pt    \font\twelvei=cmmi10 scaled 1200
 \font\twelvesy=cmsy10 scaled 1200   \font\twelveex=cmex10 scaled 1200
 \font\twelvebf=cmbx10 scaled 1200   \font\twelvesl=cmsl10 scaled 1200
 \font\twelvett=cmtt10 scaled 1200   \font\twelveit=cmti10 scaled 1200

 \skewchar\twelvei='177   \skewchar\twelvesy='60


 \def\twelvepoint{\normalbaselineskip=12.4pt
   \abovedisplayskip 12.4pt plus 3pt minus 9pt
   \belowdisplayskip 12.4pt plus 3pt minus 9pt
   \abovedisplayshortskip 0pt plus 3pt
   \belowdisplayshortskip 7.2pt plus 3pt minus 4pt
   \smallskipamount=3.6pt plus1.2pt minus1.2pt
   \medskipamount=7.2pt plus2.4pt minus2.4pt
   \bigskipamount=14.4pt plus4.8pt minus4.8pt
   \def\rm{\fam0\twelverm}          \def\it{\fam\itfam\twelveit}%
   \def\sl{\fam\slfam\twelvesl}     \def\bf{\fam\bffam\twelvebf}%
   \def\mit{\fam 1}                 \def\cal{\fam 2}%
   \def\tt{\twelvett}
   \textfont0=\twelverm   \scriptfont0=\tenrm   \scriptscriptfont0=\sevenrm
   \textfont1=\twelvei    \scriptfont1=\teni    \scriptscriptfont1=\seveni
   \textfont2=\twelvesy   \scriptfont2=\tensy   \scriptscriptfont2=\sevensy
   \textfont3=\twelveex   \scriptfont3=\twelveex  \scriptscriptfont3=\twelveex
   \textfont\itfam=\twelveit
   \textfont\slfam=\twelvesl
   \textfont\bffam=\twelvebf \scriptfont\bffam=\tenbf
   \scriptscriptfont\bffam=\sevenbf
   \normalbaselines\rm}



 \def\beginlinemode{\endmode
   \begingroup\parskip=0pt \obeylines\def\\{\par}\def\endmode{\par\endgroup}}
 \def\beginparmode{\endmode
   \begingroup \def\endmode{\par\endgroup}}
 \let\endmode=\par
 {\obeylines\gdef\
 {}}
 \def\singlespace{\baselineskip=\normalbaselineskip}
 
 \def\oneandahalfspace{\baselineskip=\normalbaselineskip
   \multiply\baselineskip by 3 \divide\baselineskip by 2}
 \def\doublespace{\baselineskip=\normalbaselineskip \multiply\baselineskip by
2}
 
 \newcount\firstpageno
 \firstpageno=1
 \footline={\hfil\twelverm\folio\hfil}
 \def\toppageno{\global\footline={\hfil}\global\headline
   ={\ifnum\pageno<\firstpageno{\hfil}\else{\hfil\twelverm\folio\hfil}\fi}}
 \let\rawfootnote=\footnote              
 \def\footnote#1#2{{\rm\singlespace\parindent=0pt\parskip=0pt
   \rawfootnote{#1}{#2\hfill\vrule height 0pt depth 6pt width 0pt}}}
 \def\raggedcenter{\leftskip=4em plus 12em \rightskip=\leftskip
   \parindent=0pt \parfillskip=0pt \spaceskip=.3333em \xspaceskip=.5em
   \pretolerance=9999 \tolerance=9999
   \hyphenpenalty=9999 \exhyphenpenalty=9999 }
 \def\dateline{\rightline{\ifcase\month\or
   January\or February\or March\or April\or May\or June\or
   July\or August\or September\or October\or November\or December\fi
   ,\space\number\year}}
 \def\received{\vskip 3pt plus 0.2fill
  \centerline{\sl (Received\space\ifcase\month\or
   January\or February\or March\or April\or May\or June\or
   July\or August\or September\or October\or November\or December\fi
   \qquad, \number\year)}}


 \hsize=16.0 true cm
 \hoffset=0.0 truein    
 \vsize=24 true cm
 \voffset=0.0 truein   
 \parskip=\medskipamount
 \def\\{\cr}
 \twelvepoint            
 \oneandahalfspace         
 \def\sectionbreak{ \penalty-200 \vskip12pt plus4pt minus6pt }
 \def\refbreak{ \penalty-300 \vskip24pt plus4pt minus6pt }
 \overfullrule=0pt       





 \def\title                      
   {\null\vskip 3pt plus 0.2fill
    \beginlinemode \doublespace \raggedcenter \bf}

 \def\author                     
   {\vskip 3pt plus 0.2fill \beginlinemode
    \singlespace \raggedcenter}

 \def\affil                      
   {\vskip 3pt plus 0.1fill \beginlinemode
    \oneandahalfspace \raggedcenter \sl}

 \def\abstract                   
   {\vskip 3pt plus 0.3fill \beginparmode
    \oneandahalfspace
    \centerline{\bf Abstract}

                                 }

 \def\endtopmatter               
   {\endpage                     
    \body}

 \def\body                       
   {\beginparmode}               

 \def\head#1{                    
   \goodbreak\vskip 0.5truein    
   {\write16{#1}
    \write2{\string\goodbreak\string\vskip0.25truein
    \string\noindent{#1}
    \string\hfill\ifnum\pageno<0{(\folio)}\else
    \folio\fi\string
    \nobreak\string\vskip0.25truein\string\nobreak}
    \raggedcenter \uppercase{#1}\par}
    \nobreak\vskip 0.25truein\nobreak}

 \def\subhead#1{                 
   \vskip 0.25truein             
   {\write2{\string\par{#1}\string\dotfill
    \ifnum\pageno<0{(\folio)}\else\folio\fi}
    \raggedcenter #1 \par}
    \nobreak\vskip 0.25truein\nobreak}

 \def\beneathrel#1\under#2{\mathrel{\mathop{#2}\limits_{#1}}}

 \def\refto#1{$[#1]$}           

 \gdef\refis#1{\item{#1.\ }}                     

 \def\endreferences{\body}

 \def\endpage                    
   {\vfill\eject}

 \def\endpaper                   
   {\endmode\vfill\supereject}


 \def\heading                            
   {\vskip 0.5truein plus 0.1truein      
    \beginparmode \def\\{\par} \parskip=0pt \singlespace \raggedcenter}

 \def\subheading                         
   {\vskip 0.25truein plus 0.1truein     
    \beginlinemode \singlespace \parskip=0pt \def\\{\par}\raggedcenter}

 \def\tag#1$${\eqno(#1)$$}

 \def\align#1$${\eqalign{#1}$$}

 \def\aligntag#1$${\gdef\tag##1\\{&(##1)\cr}\eqalignno{#1\\}$$
   \gdef\tag##1$${\eqno(##1)$$}}

 \def\endaligntag{}

 \def\overset#1\to#2{{\mathop{#2}^{#1}}}
 \def\underset#1\to#2{{\mathop{#2}_{#1}}}


 \def\ref#1{Ref.~#1}                     
 \def\Ref#1{Ref.~#1}                     
 \def\[#1]{[\cite{#1}]}
 \def\cite#1{{#1}}
 \def\(#1){(\call{#1})}
 \def\call#1{{#1}}
 \def\taghead#1{}
 \def\frac#1#2{{#1 \over #2}}

 \def\12{{1\over2}}

 \def\etal{{\it et al.\ }}

 \def\sla{\raise.15ex\hbox{$/$}\kern-.57em}
 \def\leaderfill{\leaders\hbox to 1em{\hss.\hss}\hfill}
 \def\twiddle{\lower.9ex\rlap{$\kern-.1em\scriptstyle\sim$}}
 \def\bigtwiddle{\lower1.ex\rlap{$\sim$}}
 \def\gtwid{\mathrel{\raise.3ex\hbox{$>$\kern-.75em\lower1ex\hbox{$\sim$}}}}
 \def\ltwid{\mathrel{\raise.3ex\hbox{$<$\kern-.75em\lower1ex\hbox{$\sim$}}}}
 \def\square{\kern1pt\vbox{\hrule height 1.2pt\hbox{\vrule width 1.2pt\hskip
3pt
    \vbox{\vskip 6pt}\hskip 3pt\vrule width 0.6pt}\hrule height 0.6pt}\kern1pt}
 \def\tdot#1{\mathord{\mathop{#1}\limits^{\kern2pt\ldots}}}
 
 \def\pmb#1{\setbox0=\hbox{#1}%
   \kern-.025em\copy0\kern-\wd0
   \kern  .05em\copy0\kern-\wd0
   \kern-.025em\raise.0433em\box0 }

 \openout 2=CONTENTS    
 \newcount\subsectnum \subsectnum=0
 \global\subsectnum = 0

 \outer\def\subsection#1{
 \bigskip\noindent\sectionbreak
 \write 2{\string\contsubhead {#1}{\the\pageno}}
 \global\advance\subsectnum by 1
        \noindent{\bf \sectname.\the\subsectnum }
        \quad{ {\bf #1 } }
       \medskip }

 \outer\def\secthead#1#2{\subhead{#1}\head{#2}
 \write 2{\string\conthead{\sectname}{#1}{#2}{\the\pageno}}}

 \def\references                 
   {\refbreak\head{References}
 \write 2{\string\contsubhead{\sectname}{References}{\the\pageno}}
    \beginparmode                  
    \frenchspacing \parindent=0pt \leftskip=1truecm
    \parskip=8pt plus 3pt \everypar{\hangindent=\parindent}}

\catcode`@=11
\newcount\r@fcount \r@fcount=0
\newcount\r@fcurr
\immediate\newwrite\reffile
\newif\ifr@ffile\r@ffilefalse
\def\w@rnwrite#1{\ifr@ffile\immediate\write\reffile{#1}\fi\message{#1}}

\def\writer@f#1>>{}
\def\referencefile{
\r@ffiletrue\immediate\openout\reffile=\jobname.ref
  \def\writer@f##1>>{\ifr@ffile\immediate\write\reffile%
    {\noexpand\refis{##1} = \csname r@fnum##1\endcsname = %
     \expandafter\expandafter\expandafter\strip@t\expandafter%
     \meaning\csname r@ftext\csname r@fnum##1\endcsname\endcsname}\fi}%
  \def\strip@t##1>>{}}

\def\citeall#1{\xdef#1##1{#1{\noexpand\cite{##1}}}}
\def\cite#1{\each@rg\citer@nge{#1}} 

\def\each@rg#1#2{{\let\thecsname=#1\expandafter\first@rg#2,\end,}}
\def\first@rg#1,{\thecsname{#1}\apply@rg} 
\def\apply@rg#1,{\ifx\end#1\let\next=\relax
\else,\thecsname{#1}\let\next=\apply@rg\fi\next}

\def\citer@nge#1{\citedor@nge#1-\end-} 
\def\citer@ngeat#1\end-{#1}
\def\citedor@nge#1-#2-{\ifx\end#2\r@featspace#1 
  \else\citel@@p{#1}{#2}\citer@ngeat\fi} 
\def\citel@@p#1#2{\ifnum#1>#2{\errmessage{Reference range #1-#2\space is bad.}%
    \errhelp{If you cite a series of references by the notation M-N, then M and
    N must be integers, and N must be greater than or equal to M.}}\else%
 {\count0=#1\count1=#2\advance\count1
by1\relax\expandafter\r@fcite\the\count0,%
  \loop\advance\count0 by1\relax
    \ifnum\count0<\count1,\expandafter\r@fcite\the\count0,%
  \repeat}\fi}

\def\r@featspace#1#2 {\r@fcite#1#2,} 
\def\r@fcite#1,{\ifuncit@d{#1}
    \newr@f{#1}%
    \expandafter\gdef\csname r@ftext\number\r@fcount\endcsname%
                     {\message{Reference #1 to be supplied.}%
                      \writer@f#1>>#1 to be supplied.\par}%
 \fi%
 \csname r@fnum#1\endcsname}
\def\ifuncit@d#1{\expandafter\ifx\csname r@fnum#1\endcsname\relax}%
\def\newr@f#1{\global\advance\r@fcount by1%
    \expandafter\xdef\csname r@fnum#1\endcsname{\number\r@fcount}}

\let\r@fis=\refis   
\def\refis#1#2#3\par{\ifuncit@d{#1}
   \newr@f{#1}%
   \w@rnwrite{Reference #1=\number\r@fcount\space is not cited up to now.}\fi%
  \expandafter\gdef\csname r@ftext\csname r@fnum#1\endcsname\endcsname%
  {\writer@f#1>>#2#3\par}}

\def\ignoreuncited{
   \def\refis##1##2##3\par{\ifuncit@d{##1}%
     \else\expandafter\gdef\csname r@ftext\csname
r@fnum##1\endcsname\endcsname%
     {\writer@f##1>>##2##3\par}\fi}}

\def\r@ferr{\endreferences\errmessage{I was expecting to see
\noexpand\endreferences before now;  I have inserted it here.}}
\let\r@ferences=\references
\def\references{\r@ferences\def\endmode{\r@ferr\par\endgroup}}

\let\endr@ferences=\endreferences
\def\endreferences{\r@fcurr=0
  {\loop\ifnum\r@fcurr<\r@fcount
    \advance\r@fcurr by 1\relax\expandafter\r@fis\expandafter{\number\r@fcurr}%
    \csname r@ftext\number\r@fcurr\endcsname%
  \repeat}\gdef\r@ferr{}\endr@ferences}


\let\r@fend=\endpaper\gdef\endpaper{\ifr@ffile
\immediate\write16{Cross References written on []\jobname.REF.}\fi\r@fend}

\catcode`@=12

\citeall\refto  
\citeall\ref  %
\citeall\Ref  %

\ignoreuncited

\catcode`@=11
\newcount\tagnumber\tagnumber=0

\immediate\newwrite\eqnfile
\newif\if@qnfile\@qnfilefalse
\def\write@qn#1{}
\def\writenew@qn#1{}
\def\w@rnwrite#1{\write@qn{#1}\message{#1}}
\def\@rrwrite#1{\write@qn{#1}\errmessage{#1}}

\def\taghead#1{\gdef\t@ghead{#1}\global\tagnumber=0}
\def\t@ghead{}

\expandafter\def\csname @qnnum-3\endcsname
  {{\t@ghead\advance\tagnumber by -3\relax\number\tagnumber}}
\expandafter\def\csname @qnnum-2\endcsname
  {{\t@ghead\advance\tagnumber by -2\relax\number\tagnumber}}
\expandafter\def\csname @qnnum-1\endcsname
  {{\t@ghead\advance\tagnumber by -1\relax\number\tagnumber}}
\expandafter\def\csname @qnnum0\endcsname
  {\t@ghead\number\tagnumber}
\expandafter\def\csname @qnnum+1\endcsname
  {{\t@ghead\advance\tagnumber by 1\relax\number\tagnumber}}
\expandafter\def\csname @qnnum+2\endcsname
  {{\t@ghead\advance\tagnumber by 2\relax\number\tagnumber}}
\expandafter\def\csname @qnnum+3\endcsname
  {{\t@ghead\advance\tagnumber by 3\relax\number\tagnumber}}

\def\equationfile{%
  \@qnfiletrue\immediate\openout\eqnfile=\jobname.eqn%
  \def\write@qn##1{\if@qnfile\immediate\write\eqnfile{##1}\fi}
  \def\writenew@qn##1{\if@qnfile\immediate\write\eqnfile
    {\noexpand\tag{##1} = (\t@ghead\number\tagnumber)}\fi}
}

\def\callall#1{\xdef#1##1{#1{\noexpand\call{##1}}}}
\def\call#1{\each@rg\callr@nge{#1}}

\def\each@rg#1#2{{\let\thecsname=#1\expandafter\first@rg#2,\end,}}
\def\first@rg#1,{\thecsname{#1}\apply@rg}
\def\apply@rg#1,{\ifx\end#1\let\next=\relax%
\else,\thecsname{#1}\let\next=\apply@rg\fi\next}

\def\callr@nge#1{\calldor@nge#1-\end-}
\def\callr@ngeat#1\end-{#1}
\def\calldor@nge#1-#2-{\ifx\end#2\@qneatspace#1 %
  \else\calll@@p{#1}{#2}\callr@ngeat\fi}
\def\calll@@p#1#2{\ifnum#1>#2{\@rrwrite{Equation range #1-#2\space is bad.}
\errhelp{If you call a series of equations by the notation M-N, then M and
N must be integers, and N must be greater than or equal to M.}}\else%
 {\count0=#1\count1=#2\advance\count1
by1\relax\expandafter\@qncall\the\count0,%
  \loop\advance\count0 by1\relax%
    \ifnum\count0<\count1,\expandafter\@qncall\the\count0,%
  \repeat}\fi}

\def\@qneatspace#1#2 {\@qncall#1#2,}
\def\@qncall#1,{\ifunc@lled{#1}{\def\next{#1}\ifx\next\empty\else
  \w@rnwrite{Equation number \noexpand\(>>#1<<) has not been defined yet.}
  >>#1<<\fi}\else\csname @qnnum#1\endcsname\fi}

\let\eqnono=\eqno
\def\eqno(#1){\tag#1}
\def\tag#1$${\eqnono(\displayt@g#1 )$$}

\def\aligntag#1\endaligntag
  $${\gdef\tag##1\\{&(##1 )\cr}\eqalignno{#1\\}$$
  \gdef\tag##1$${\eqnono(\displayt@g##1 )$$}}

\def\eqalignno#1{\displ@y \tabskip\centering
  \halign to\displaywidth{\hfil$\displaystyle{##}$\tabskip\z@skip
    &$\displaystyle{{}##}$\hfil\tabskip\centering
    &\llap{$\displayt@gpar##$}\tabskip\z@skip\crcr
    #1\crcr}}

\def\displayt@gpar(#1){(\displayt@g#1 )}

\def\displayt@g#1 {\rm\ifunc@lled{#1}\global\advance\tagnumber by1
        {\def\next{#1}\ifx\next\empty\else\expandafter
        \xdef\csname @qnnum#1\endcsname{\t@ghead\number\tagnumber}\fi}%
  \writenew@qn{#1}\t@ghead\number\tagnumber\else
        {\edef\next{\t@ghead\number\tagnumber}%
        \expandafter\ifx\csname @qnnum#1\endcsname\next\else
        \w@rnwrite{Equation \noexpand\tag{#1} is a duplicate number.}\fi}%
  \csname @qnnum#1\endcsname\fi}

\def\ifunc@lled#1{\expandafter\ifx\csname @qnnum#1\endcsname\relax}

\let\@qnend=\end\gdef\end{\if@qnfile
\immediate\write16{Equation numbers written on []\jobname.EQN.}\fi\@qnend}

\catcode`@=12

\catcode`@=11 
\def\lsim{\mathrel{\mathpalette\@versim<}}
\def\gsim{\mathrel{\mathpalette\@versim>}}
\def\@versim#1#2{\lower0.2ex\vbox{\baselineskip\z@skip\lineskip\z@skip
  \lineskiplimit\z@\ialign{$\m@th#1\hfil##\hfil$\crcr#2\crcr\sim\crcr}}}
\catcode`@=12 

\def\pagenumber{\pageno=1\global\footline={\hfil\twelverm\folio\hfil}}

\def\LEPII/{LEP I\hskip -1pt I}
\def\GeV{{\rm\ GeV}}
\def\MeV{{\rm\ MeV}}
\def\TeV{{\rm\ TeV}}
\def\pb{{\rm\ pb}}
\def\alphas{\alpha_{\scriptscriptstyle S}}
\font\circle=circle10
\font\mii=cmmi10
\def\trcorner{\hbox{\raise 2.7pt\hbox{\circle\char10}\kern -10pt}}
\def\barrow{\trcorner\to}

\def\sectname{}
\nopagenumbers
\hfill\vbox{\rm\hbox{DTP/93/2}
\hbox{February 1993}}
\title
Calculation of the contributions to the signals and backgrounds for
intermediate mass Higgs detection at the LHC and SSC from the
$q \bar q$ initial state
\bigskip
\author
D.J. Summers
\affil
Centre for Particle Theory,
University of Durham,
Durham DH1 3LE, England.
\bigskip
\abstract
We calculate the subprocess $q \bar q\to t \bar t H$ which contributes
to the signal for detection of a light `intermediate mass' Higgs boson
at the LHC and SSC in the isolated lepton and two isolated photons
mode. This enhances the $gg\to t \bar t H$ signal by about 10\%
at the SSC and 25\% at the LHC. We also calculate the
$q \bar q \to t \bar t \gamma\gamma$ irreducible background subprocess to
this detection mode which enhances the $gg \to t \bar t \gamma\gamma$
irreducible background. This background is found to be quite
substantial due to the effect of radiation from the initial state
quarks as well as the final state $t \bar t$ pair; at the SSC the
$q \bar q$ initial state background enhances the $gg$ initial state
background by between 20\% for $m_t=100\GeV$ and 35\% for
$m_t=180\GeV$, and at the LHC by between 50\% for $m_t=100\GeV$ and
120\% for $m_t=180\GeV$. However with a 3\% mass resolution on the
$\gamma\gamma$ mass the extra $q \bar q$ background events are
typically smaller than the extra $q \bar q$ signal events, and the
signal remains observable.
\endpage
\body
\pagenumber

Over the past two decades the Standard Model has been spectacularly
verified with the discovery of $W$ and $Z$ bosons, however the mass
generation sector of the Standard Model is very poorly constrained.
In the Standard Model a complex doublet scalar field generates the
mass of the $W$ and $Z$ via the Higgs mechanism, and the fermion masses
via the Yukawa mechanism; after the generation of vector boson masses
there remains one neutral $CP$ even physical scalar in the Standard
Model, the Higgs boson. This Higgs boson has unknown mass, however
there are theoretical biases for believing the Higgs mass, $M_H \lsim 1
\TeV$\refto{HHG}.  Recent negative searches from LEPI have show that
$M_H > 57 \GeV$\refto{57}. This leaves the mass range $57 \GeV < M_H
\lsim 1 \TeV$ in which we expect one (or possibly more, in extensions
to the Standard Model) Higgs boson to exist. It is a major aim of the
next round of accelerators to cover this mass range and either detect,
or rule out, the Higgs boson. \LEPII/ will be able to detect a Higgs
with $M_H \lsim 80 \GeV$ \refto{Wu}, while the LHC and the SSC probe
the heavier Higgs sector through the $H \to ZZ \to 4l$ decay mode for
$M_H > 2 M_Z$; and via $H \to ZZ^* \to 4l$ for $2 M_Z > M_H \gsim 130
\GeV$. For Higgs with $80 \GeV \lsim M_H \lsim 130 \GeV$ the most
promising detection mode is to detect the Higgs in association with
another heavy particle at high energy hadron colliders.
The most promising signals come from\refto{ttH,WH},
$$\eqalign{ pp \to &t \bar t H \to \gamma \gamma \cr
\noalign{\vskip -7pt}
                &\hbox{\hskip -1pt\raise 9pt\hbox{\mii\char94}\kern -5pt}
\barrow lX \cr}
\tag ttH
$$
$$\eqalign{ pp \to &W H \to \gamma \gamma \cr
\noalign{\vskip -9pt}
                  &\hbox{\kern 5pt}\barrow l\nu \cr}
\tag WH
$$
where we tag both signals by the presence of two isolated photons from
the Higgs decay and an isolated lepton ($l=\mu ,e$) from the associated heavy
particle decay. Even though we tag on the same signal, the two types of
event look considerably different; process \(ttH) regularly contains
$2$ or more well defined extra jets from the $t \bar t$ decay products,
whereas in process \(WH) each extra jet costs us an extra power of
$\alphas$. This means that we can consider the two processes,
\(ttH,WH), independently; and as such calculate the backgrounds
for each process separately.

If we can reject jets from looking like photons at $1$ part in $10^4$
then the major background for process \(ttH) has been show to be
\refto{ggttff,meggttff},
$$\eqalign{ pp \to &t \bar t \gamma \gamma \cr
\noalign{\vskip -7pt}
       &\hbox{\hskip -1pt\raise 9pt\hbox{\mii\char94}\kern -5pt}
\barrow lX \cr}
\qquad.\tag ppttff
$$
This has been calculated from the subprocess,
$$ gg \to t \bar t \gamma \gamma \tag ggttff$$
which, on the face of it, dominates over the subprocess,
$$ q \bar q \to t \bar t \gamma \gamma \tag qqttff$$
due to the large gluon luminosity at small Bjorken $x$ relative to the
quark luminosities, and also the colour structure of the matrix
elements. For example for just $t \bar t$ production the $q \bar q$
initial state contribution is typically 10\% at LHC energies
($\sqrt{s}=16\TeV$) and 6\% at SSC energies ($\sqrt{s}=40\TeV$) of the
$gg$ initial state contribution.

However in \(qqttff) the photons can be radiated from the initial
state quarks as well as from the final state $t \bar t$ pair. This
means that that we have far more Feynman diagrams for process
\(qqttff), so we not only have just the final state radiation, but
also initial state radiation, mixed initial-final state radiation, as
well as all the interference between all the terms. Also when photons
are radiated from the initial state the exchanged gluon in process
\(ppttff) has less energy flowing through it and hence is closer to its
mass shell; this can enhance the initial state radiation by a factor
of 3 or so. Of course the initial quarks may have charge
$\scriptscriptstyle 1 \over \scriptscriptstyle 3$ rather than the $t$
quark charge of $\scriptscriptstyle 2 \over \scriptscriptstyle 3$
which will lessen the effect of initial state radiation. All in all,
it is uncertain as to how large the $q \bar q$ background process
\(qqttff) will be relative to the $gg$ background process \(ggttff).
Here we calculate the process
\(qqttff) and also the $q \bar q$ signal process,
$$ q \bar q \to t \bar t H \tag qqttH$$
which supplements the $gg$ signal process,
$$ gg \to t \bar t H \tag ggttH$$
For the background matrix element \(qqttff) we have adapted the code
of \Ref{mmff} where the process $e^+e^- \to \mu^+ \mu^- + n \gamma$
was calculated keeping the full mass dependence of the muon. We change
the exchanged photon and $Z$ into an exchanged gluon, the electrons
into our initial state quarks and the muons into our final state $t$
quarks. We calculate the signal matrix element \(qqttH) using the
spinor techniques of \Ref{Spinor}, and introduce the mass of the $t$
quark using the massive extension to massless spinors defined in
\Ref{massive}.

We convolute the subprocess cross sections with the MRS D$0'$ parton
distributions of \Ref{MRS92p} to obtain full cross sections for the $pp$
initial state. For consistency with these parton distributions we set
$\Lambda_{\scriptscriptstyle\overline{MS}}^{\scriptscriptstyle(4)}=
230 \MeV$ we rescale this to
$\Lambda_{\scriptscriptstyle\overline{MS}}^{\scriptscriptstyle(5)}$
by forcing agreement between
$\alphas^{\scriptscriptstyle NF=4}$ and
$\alphas^{\scriptscriptstyle NF=5}$ at $Q=m_b$ and calculate the
running of $\alphas$ at the 1 loop level. We use $5$ active flavors
in the running of $\alphas$ for $Q>m_b$.

After generating the $t \bar t H$ final state we decay the Higgs
isotropically to two photons using the $BR(H \to \gamma \gamma)$ from
\Ref{brH}. Then for both the signal and the background processes we
decay the
$t$ quarks via a $V-A$ interaction to onshell $W$'s and then decay
the $W$'s again by a $V-A$ interaction. We insist that at least one of
the $W$'s decay to a lepton ($e$,$\mu$).

For comparison we have included the cross sections from the $gg$
initial state process, \(ggttH,ggttff) from \Ref{meggttff}, calculated
using the same parameters as for the $q \bar q$ initial states. We
calculate the $gg$ and $q \bar q$ initial state cross sections using
the same Monte Carlo events, this means that the Monte Carlo error
tends to cancel in the ratio of $q \bar q$ to $gg$ initial state events.

For the various electoweak parameters we use $M_Z = 91.1\GeV$,
$\sin^2 \theta_W = 0.23$, $\alpha_{em}= 1/137.04$, and $M_W = M_Z \cos
\theta_W$.
\goodbreak

We use the following cuts to simulate a detector;
$$\eqalignno{
p_T(l,\gamma )>20\ GeV \qquad&,\qquad |\eta(l,\gamma )|<2.5           \cr
\Delta R(\gamma_1,\gamma_2)>0.4 \qquad&,\qquad \Delta R(l,\gamma)>0.4
&(cuts)\cr
\Delta R(\gamma,\hbox{jet})>0.4 \qquad&,\qquad \Delta R(l,\hbox{jet})>0.4 \cr
}$$
where we associate a jet with each final state quark that we have.
$\Delta R$ is defined as $\sqrt{\Delta \phi^2 + \Delta \eta^2}$ where
$\eta$ is the rapidity and $\phi$ the azimuthal angle about the beam
direction.

We obtain the results shown in figures 1a) at the SSC
($\sqrt{s}=40\TeV$) and 1b) at the LHC ($\sqrt{s}=16\TeV$),
with 3 choices for $m_t = 100 , 140
, 180 \GeV$ where we have binned the data in $5\GeV$
$M_{\gamma\gamma}$ bins. We also show the number of events expected
for $\sqrt{s}=16\TeV$ (LHC) with integrated luminosity ${\cal L}=10^5
\pb^{-1}$ and $\sqrt{s}=40\TeV$ (SSC) with integrated luminosity
${\cal L}=10^4 \pb^{-1}$. In practice we expect a better diphoton mass
resolution, and this will lower the continuum background without
affecting the signal (recall that for a Higgs in this mass range the
Higgs width is ${\cal O}$(MeV) and as such the signal diphoton mass
distribution is narrower than we will be able to experimentally
measure.)

In tables 1a) and 1b) we show the expected number of events at the SSC
($\sqrt{s}=40\TeV$) with ${\cal L}=10^4 \pb^{-1}$ and at the LHC
($\sqrt{s}=16\TeV$) with ${\cal L}=10^5 \pb^{-1}$ with
$M_{\gamma\gamma}$ bin widths of
$3\%$ of $M_{\gamma\gamma}$.

{}From the figures and tables it is clear that for both the signal and
background the effect of the $q \bar q$ initial state is far larger at
the LHC than at the SSC. This is because we are at smaller $\sqrt{s}$
at the LHC and so for the same $\sqrt{\hat s}$ we require larger
Bjorken $x$'s and so we feel the effect of the valence quarks in the
proton more. For the signals the $q \bar q$ initial state is about
$10\%$ at the SSC, and $25\%$ at the LHC, of the $gg$ initial state;
this is about double the enhancement that we get for just $t
\bar t$ production without the $H$, this is because the extra particle
in the final state (the Higgs) forces the $\sqrt{\hat s}$ to be larger
and so forces us to larger Bjorken $x$ where we again feel the effect
of the valence quarks more. At the SSC the $q \bar q$ background is
$20\%$ of the $gg$ background for $m_t=100 \GeV$ increasing to $35\%$
for $m_t=180 \GeV$. At the LHC the $q \bar q$ background is $50\%$ of
the $gg$ background for $m_t = 100 \GeV$ increasing to $120\%$ for
$m_t=180 \GeV$. It is clear that the $q \bar q$ initial state is a
very important source of background events, especially for larger
$m_t$, although fortunately for larger $m_t$ the
$t \bar t \gamma\gamma$
background is far smaller than the $t \bar t H$ signal. For
lighter $m_t$ where the $t \bar t \gamma\gamma$ background is far more
severe the $q \bar q$ is of lesser importance -- but still
significantly enhances the background.

We now wish to address the question or whether we can decrease this
large source of background events coming from the $q \bar q$ initial
state. At first glance this seems impossible as the background is an
irreducible background; however we should not forget that the
production mechanism is considerably different from the signals
\(ggttH,qqttH) and the $gg$ initiated background \(ggttff).
In the processes \(ggttH,qqttH,ggttff) the
Higgs or photons are always radiated from the final state $t \bar t$
pair, however for the $q \bar q$ background \(qqttff) we also have
contributions from photons radiated from the initial state. In
particular this means that we have singularities when the photons
become collinear with the beam direction. Of course these
singularities are regulated by the requirements that $p_T(\gamma) > 20
\GeV$ and $|\eta| < 2.5$. This means that by varying these cuts we may
be sensitive to process \(qqttff) and hence be able to reduce its
importance.

We have checked the distributions of the signal and background processes
on $p_T$ and $|\eta|$ of the photons and found that there is almost no
difference in the $p_T$ distributions, but that the $q \bar q$
background \(qqttff) has a rapidity distribution different to
the signal and other background. We show this in figure 2 where we
have relaxed the cut on rapidity. We show the $q \bar q$
background process \(qqttff) and the $gg$ background process
\(ggttff); the signals have a similar distribution to the $gg$
background process \(ggttff) although obviously with a different
normalisation. Although there is a tendency for the photons produced
in the $q \bar q$ background process to be more collinear with the
beam direction than the $gg$ background process, it is also clear that
there is no significant enhancement factor to be gained by making the
$|\eta |$ more restrictive than in \(cuts).

\endpage
\head{Conclusions}

For detection of a Standard Model Higgs boson with
$70\GeV \lsim M_H \lsim 130\GeV$ via
the isolated lepton plus two isolated photon detection mode,  the
$t \bar t H$ signal receives a helpful few events from the $q \bar q$
initial state signal events; at the LHC this
enhances the signal by about $25\%$, and at the SSC by about $10\%$.

However the main irreducible background process \(ppttff) also
receives large contributions from the $q \bar q$ initial state
\(qqttff) relative to the $gg$ initial state \(ggttff), indeed at the
LHC for heavy top this $q \bar q$ initial state background
dominates the $gg$ initial state background.

If we can achieve a modest $3\%$ of $M_{\gamma\gamma}$ mass resolution
on $M_{\gamma\gamma}$ then the isolated lepton plus two isolated
photon detection mode is still observable above the main irreducible
background \(ppttff), the additional source of background events from
process \(qqttff) is typically smaller than the extra signal
events from process \(qqttH).

This can be achieved with the standard luminosity of
${\cal L}=10^4 \pb^{-1}$ at the SSC ($\sqrt{s}=40\TeV$), however
at the LHC ($\sqrt{s}=16\TeV$) we require higher than the standard
luminosity of ${\cal L}=10^4 \pb^{-1}$, ${\cal L}=5\times 10^4
\pb^{-1}$ will be sufficient.

This detection mode can be extended up to Higgs masses of about
$150\GeV$, however for higher Higgs mass the branching ratio into two
photons drops rapidly as the width for $H \to WW^*$ grows rapidly. It
can also be extended down to lower current LEPI mass limit of
$60\GeV$, however for lower mass Higgs the $H\to\gamma\gamma$
branching ratio drops off and the $t\bar t \gamma\gamma$ background
grows very rapidly. It is useful to extend this detection mode as far
as possible because it tests the coupling of the Higgs to both $W$
bosons (and hence vector boson mass generation) and $t$ quarks (and
hence fermion mass generation), because we can distinguish the two
signals \(ttH) and \(WH).

\head{Acknowledgments}

I would like to thank James Stirling for useful discussions. I would
also like to thank SERC for financial support in the form of a
research studentship.

\endpage

\references

\refis{HHG} See for example : J.F.Gunion, G.L.Kane, H.E.Haber, and S.Dawson,
{\it The Higgs Hunter's Guide}, Addison Wesley, (1990).

\refis{WH} R.Kleiss, Z.Kunszt, W.J.Stirling, {\it Phys.Lett.} {\bf B253}
(1991)269.

\refis{ngamma} R.Kleiss and W.J.Stirling, {\it Phys.Lett.} {\bf 179B} (1986)
159.

\refis{strfns} J.Kwiecinski, A.D.Martin, W.J.Stirling and R.G.Roberts,
{\it Phys.Rev.} {\bf D42} (1990) 3645.

\refis{MRS92p} A.D. Martin, R.G. Roberts, and W.J. Stirling, Durham
University preprint, DTP/92/80.

\refis{Peterson} C.Peterson, D.Schlatter, I.Schmitt and P.M.Zerwas
{\it Phys.Rev.} {\bf D27} (1983)105.

\refis{1/7} C.Seez \etal , Proceedings of the Large Hadron Collider Workshop,
Aachen, CERN 90-10 Vol. II (1990) 474.

\refis{massive} R. Kleiss and W.J. Stirling, {\it Phys.Lett.} {\bf 179B} (1986)
159. \hfil\break
F.A. Berends, P.H. Daverveldt and R. Kleiss, {\it Nucl. Phys.} {\bf
B253} (1985) 411.

\refis{Spinor} R. Kleiss and W.J. Stirling, {\it Nucl. Phys.}
{\bf B262} (1985) 235.

\refis{spinor2} J.F. Gunion and Z. Kunszt, {\it Phys.Lett.} {\bf 161B} (1985)
333.

\refis{L3hardphot} The L3 Collaboration, CERN preprint, CERN-PPE/92-152

\refis{Jamesmmnf} W.J. Stirling {\it Phys.Lett.} {\bf 271B} (1991)
261.

\refis{soft} R. Kleiss \etal , `$Z$ Physics at LEP1', CERN Yellow
Report 89-08 (1989) vol.3, page 1. \hfil\break
S. Jadach and B.F.L Ward, {\it Phys. Lett.} {\bf 274B} (1992) 470.

\refis{ZLS} F.A. Berends \etal , `$Z$ Physics at LEP1', CERN Yellow
Report 89-08 (1989), vol.1, page 89.

\refis{2loop} F.A. Berends, G. Burgers and W.L. van Neerven, {\it
Nucl. Phys.} {\bf B297} (1988) 429; E {\it Nucl. Phys.} {\bf B304}
(1988) 921.

\refis{calcutmm2f} F.A. Berends \etal , {\it Nucl. Phys.} {\bf B264}
(1986) 243.

\refis{ttH} W.J.Marciano and F.E.Paige, {\it Phys.Rev.Lett.}
{\bf 66} (1991) 2433.\hfil\break
J.F.Gunion, {\it Phys.Lett.} {\bf B261} (1991) 510.

\refis{Ral} See for example : W.J.Stirling {\it Topics in Modern
Phenomenology}, Lecture Notes - HEP Summer School, RAL (1987)

\refis{Wu} S.L.Wu \etal ECFA workshop on LEP200, eds A.B\"ohm and W.Hoogland,
Vol II (1987) 312.

\refis{Snow} Proceedings of 1988 Snowmass Workshop on  ``Physics in the 1990's
'', ed. S. Jensen, World Scientific (1989).

\refis{Cern} Proceedings of the Workshop on Physics at Future Accelerators La
Thuile, ed. J. Mulvey, CERN Yellow Report 87-07 (1987).

\refis{Spinor} R. Kleiss and W.J. Stirling, {\it Nucl. Phys.} {\bf B262}
(1985)235.

\refis{spinor2} J.F.Gunion and Z.Kunszt, {\it Phys.Lett.} {\bf 161B} (1985)
333.

\refis{James1} Z.Kunszt and W.J.Stirling, {\it Phys. Lett.} {\bf 242B} (1990)
507.

\refis{James2} R.Kleiss, Z.Kunszt, and W.J.Stirling {\it Phys.Lett.}
{\bf 253B} (1991) 269.

\refis{Nick} N.Brown  {\it Z.Phys.} {\bf C49} (1991) 657.

\refis{qcdwh} T.Han and S.Willenbrock, Fermilab preprint FERMILAB-PUB-91-70-T,
\hbox{BNL-45990} (1991).

\refis{brH} Z.Kunszt and W.J.Stirling, Aachen ECFA Workshop (1990) 428.

\refis{same1} Z.Kunszt, Z.Trocsanyi and W.J.Stirling, Durham University
preprint DTP-91-40, ETH-TH-91-17 (1991).

\refis{ttH} W.J.Marciano and F.E.Paige, {\it Phys. Rev. Lett.}
{\bf 66} (1991) 2433.\hfil\break
J.F.Gunion, {\it Phys. Lett.} {\bf B261} (1991) 510.

\refis{ggttff} Z.Kunszt, Z.Trocsanyi and W.J.Stirling, {\it Phys.
Lett.} {\bf B271} (1991) 247.\hfil\break
A.Ballestrero and E.Maina, {\it Phys. Lett.} {\bf B268} (1991)
437.

\refis{meggttff} D.J.Summers, {\it Phys.Lett.} {\bf B277} (1992) 366.

\refis{same2} A.Ballestrero and E.Maina, Turin University preprint
DFTT-29-91 (1991).

\refis{48} ALEPH Collaboration D.Decamp \etal preprint CERN-PPE/91-19,
Contribution to the Aspen, La Thuile and Moriond Conferences, winter 1991.

\refis{57} M.Davier, Plenary talk, Joint Lepton-Photon Symposium and EPS
Conference, Geneva (1991).

\refis{hmass} J.Ellis, G.Ridolfi and F.Zwirner, {\it  Phys.Lett.}
{\bf 257B} (1991) 83.\hfil\break
H.E.Haber and  R.Hempfling, {\it Phys.Rev.Lett.}
{\bf 66} (1991) 1815.\hfil\break
Y.Okada, M.Yamaguchi and T. Yanagida, Tohoku University
preprint TU-360 (1990).

\refis{mmff} D.J.Summers, Durham University preprint, DTP/92/76

\endreferences

\endpage
\nopagenumbers
\head{Figure Captions}
\parindent=50pt
\item{Figure 1. }
The differential cross sections ($d\sigma/dM_{\gamma\gamma}$ (fb/5\GeV))
for the process $gg \to t \bar t \gamma \gamma$, shown with light
lines, and for $gg + q \bar q \to t \bar t \gamma \gamma$, shown with
thick lines at the a) the SSC with $\sqrt{s}=40\TeV$ and b) the LHC
with $\sqrt{s}=16\TeV$. Shown for three values of $m_t = 100,140,180
\GeV$. Also shown are the expected numbers of events for the standard
luminosity SSC (${\cal L}=10^4 \pb^{-1}$) and the high luminosity LHC
(${\cal L}=10^5 \pb^{-1}$). Superimposed are the cross sections for the
processes $gg \to t \bar t H$ (light lines) and
$gg + q \bar q \to t \bar t H$ (thick lines) for
three values of $M_H = 70,100,130 \GeV$.
\item{Figure 2. }
The differential cross sections ($d\sigma/d|\eta |$ (fb/0.25)) for the
processes $q \bar q \to t \bar t \gamma \gamma$ (solid) and
$gg \to t \bar t \gamma \gamma$ (dashed) for $m_t=140\GeV$.
%
%
\head{Tables}
\item{Table 1. }
The numbers of events for the signal \(ggttH,qqttH) and background
\(ggttff,qqttff) process at \break a) the SSC with ${\cal L}=10^4 \pb^{-1}$
and b) the LHC with ${\cal L}=10^5 \pb^{-1}$. For the background
events we assume a $3\%$ of $M_{\gamma\gamma}$ bin width in
$M_{\gamma\gamma}$ about $M_H$.
\smallskip
\vbox{

\centerline{\vbox{\offinterlineskip
\hrule
\halign{\vrule#&\strut\quad\hfil$#$\hfil\quad
&\hfil$#$\hfil\quad&\vrule#\hskip 1pt\vrule
&\quad\hfil$#$\hfil\quad&\vrule#
&\quad\hfil$#$\hfil\quad&\vrule#
&\quad\hfil$#$\hfil\quad&\vrule#
&\quad\hfil$#$\hfil\quad&\vrule#\cr
height 2pt&\omit&\omit&&\multispan7&\cr
&M_H&m_t&& \multispan7 \hfil $\sqrt{s} = 40\TeV$ \hfil&\cr
height 1pt&\omit&\omit&&\multispan7&\cr
&\omit&\omit&\multispan8\hrulefill&\cr
height 2pt&\omit&\omit&&\omit&&\omit&&\omit&&\omit&\cr
& ({\rm GeV}) & ({\rm GeV}) && gg \to t \bar t H && q \bar q \to t \bar t H
&& gg \to t \bar t \gamma \gamma && q \bar q \to t \bar t \gamma\gamma &\cr
height 2pt&\omit&\omit&&\omit&&\omit&&\omit&&\omit&\cr
\noalign{\hrule}
height 1pt&\omit&\omit&&\omit&&\omit&&\omit&&\omit&\cr
\noalign{\hrule}
height 2pt&\omit&\omit&&\omit&&\omit&&\omit&&\omit&\cr
   &     & 100 && 9.0  && 0.8  && 4.1   && 0.8  &\cr
   & 70  & 140 && 11.4 && 1.0  && 1.6   && 0.5  &\cr
   &     & 180 && 12.1 && 1.2  && 0.8   && 0.3  &\cr
\noalign{\hrule}
height 2pt&\omit&\omit&&\omit&&\omit&&\omit&&\omit&\cr
   &     & 100 && 10.7 && 1.0  && 4.4  && 0.9  &\cr
   & 100 & 140 && 14.2 && 1.5  && 1.8  && 0.6  &\cr
   &     & 180 && 16.3 && 1.9  && 0.9  && 0.3  &\cr
\noalign{\hrule}
height 2pt&\omit&\omit&&\omit&&\omit&&\omit&&\omit&\cr
   &     & 100 && 9.6  && 0.9  && 3.9  && 0.8  &\cr
   & 130 & 140 && 11.8 && 1.4  && 1.6  && 0.5  &\cr
   &     & 180 && 13.9 && 1.9  && 0.8  && 0.3  &\cr
}\hrule}}}
\smallskip
\vbox{

\centerline{\vbox{\offinterlineskip
\hrule
\halign{\vrule#&\strut\quad\hfil$#$\hfil\quad&\hfil$#$\hfil\quad&
\vrule#\hskip 1pt\vrule
&\quad\hfil$#$\hfil\quad&\vrule#
&\quad\hfil$#$\hfil\quad&\vrule#
&\quad\hfil$#$\hfil\quad&\vrule#
&\quad\hfil$#$\hfil\quad&\vrule#\cr
height 2pt&\omit&\omit&&\multispan7&\cr
& M_H\hfil & m_t\hfil && \multispan7 \hfil $\sqrt{s} = 16\TeV$ \hfil&\cr
height 1pt&\omit&\omit&&\multispan7&\cr
&\omit&\omit&\multispan8\hrulefill&\cr
height 2pt&\omit&\omit&&\omit&&\omit&&\omit&&\omit&\cr
& ({\rm GeV}) & ({\rm GeV}) && gg \to t \bar t H && q \bar q \to t \bar t H
&& gg \to t \bar t \gamma \gamma && q \bar q \to t \bar t \gamma\gamma &\cr
height 2pt&\omit&\omit&&\omit&&\omit&&\omit&&\omit&\cr
\noalign{\hrule}
height 1pt&\omit&\omit&&\omit&&\omit&&\omit&&\omit&\cr
\noalign{\hrule}
height 2pt&\omit&\omit&&\omit&&\omit&&\omit&&\omit&\cr
   &     & 100 && 17.2 &&  3.0 && 6.8  && 3.5  &\cr
   & 70  & 140 && 19.2 &&  3.8 && 2.7  && 2.0  &\cr
   &     & 180 && 18.3 &&  4.2 && 0.9  && 1.1  &\cr
\noalign{\hrule}
height 2pt&\omit&\omit&&\omit&&\omit&&\omit&&\omit&\cr
   &     & 100 && 18.6 &&  3.9 && 7.8  && 3.7  &\cr
   & 100 & 140 && 22.4 &&  5.5 && 2.6  && 2.0  &\cr
   &     & 180 && 23.2 &&  6.4 && 1.0  && 1.2  &\cr
\noalign{\hrule}
height 2pt&\omit&\omit&&\omit&&\omit&&\omit&&\omit&\cr
   &     & 100 && 15.2 &&  3.2 && 6.1  && 3.0  &\cr
   & 130 & 140 && 16.9 &&  4.9 && 2.0  && 1.8  &\cr
   &     & 180 && 18.4 &&  6.0 && 0.9  && 1.1  &\cr
}\hrule}}}
\end